\documentclass[useAMS,usenatbib]{mn2e}
\usepackage{epsfig}
\usepackage{multirow}
\usepackage{aas_macros}     
\usepackage{amsfonts}
\title[Velocity and mass bias in the distribution of dark matter halos]
{ 
Velocity and mass bias in the distribution of dark matter halos
}
\author[Jennings et al.  ]
{ Elise Jennings$^{1}$\thanks{E-mail: ejennings@kicp.uchicago.edu}, Carlton M. Baugh$^{2}$, Dylan Hatt$^{3}$\\
$^{1}$ The Kavli Institute for Cosmological Physics and Enrico Fermi Institute, University of Chicago, \\
5640 South Ellis Avenue, Chicago, IL 60637, U. S.\\
$^{2}$Institute for Computational Cosmology, Dept. of Physics, Durham University, South Road, Durham DH1 3LE, UK.\\
$^{3}$Department of Astronomy \& Astrophysics, University of Chicago, Chicago, IL 60637\\
}

\begin{document}

\date{}

\pagerange{\pageref{firstpage}--\pageref{lastpage}} 

\maketitle

\label{firstpage}

\begin{abstract}
The non-linear, scale-dependent bias in the mass distribution of galaxies
 and the underlying dark matter is a key systematic affecting the
extraction of cosmological parameters
from galaxy clustering.
Using 95 million halos from the Millennium-XXL N-body simulation, 
we find that the mass bias is scale independent {\it only} for $k<0.1 h{\rm Mpc}^{-1}$ 
today ($z=0$) and for $k<0.2 h{\rm Mpc}^{-1}$ at $z=0.7$. We test analytic halo bias 
models against our simulation measurements and find that the model of \citet{2005ApJ...631...41T} 
is accurate to better then 5\% at $z=0$. However, the simulation results are better fit by 
an ellipsoidal collapse model at $z=0.7$.
We highlight, for the first time, 
another potentially serious systematic 
due to a sampling bias in the halo velocity divergence power spectra which will affect
the comparison between observations and any redshift space distortion model which assumes dark matter velocity statistics with no velocity bias.
By measuring the velocity divergence power spectra for different sized halo samples,  
we find that there is a significant bias which increases with decreasing number density. 
This bias is approximately 20\% at $k=0.1h$Mpc$^{-1}$ for a halo sample of number density 
$\bar{n} = 10^{-3} (h/$Mpc$)^3$ at both $z=0$ and $z=0.7$ for the velocity divergence auto 
power spectrum. 
Given the importance of redshift space distortions as a probe of dark energy
and the on-going major effort to advance
 models for the clustering signal in redshift space,
our results show this velocity bias introduces another systematic, alongside scale-dependent 
halo mass bias, which cannot be neglected.
\end{abstract}


\section{Introduction}

Current and upcoming galaxy surveys such as BOSS \citep{2007AAS...21113229S}, DES \citep{2013AAS...22133501F}, DESI \citep{2013arXiv1308.0847L}, LSST \citep{2008arXiv0805.2366I} and {\sc euclid} \citep{EUCLID}
will require extremely accurate theoretical predictions to match 
the precise observations of large-scale structure in our Universe. 
Cosmological N-body simulations which combine high resolution and large 
volume have the statistical power to play a key role in guiding the 
development of accurate theoretical models which will advance our understanding of the hierarchical growth of structure, galaxy formation and the properties of dark energy. 
A large uncertainty in extracting cosmological information from observations is the bias between galaxies or dark matter halos and the underlying dark matter 
distribution. Using the {\it Millennium-XXL} ({\sc mxxl}) simulation  we examine both the halo mass and velocity bias for different mass bins and compare with theoretical predictions. To our knowledge this is the first time that 
the velocity divergence power spectra have been presented for halos of different masses measured from N-body simulations.
Accurate models for the mass and velocity bias of halos are 
extremely important in theoretical predictions 
for redshift space distortions which are a major cosmological probe in
the Dark Energy Task Force stage IV experiments \citep{2006astro.ph..9591A}.

Dark matter halos form at high fluctuation peaks in the matter distribution and represent a biased tracer of the dark matter \citep[e.g.][]{1986ApJ...304...15B}.
As a consequence,
extracting cosmological parameters from clustering statistics requires an accurate model for this bias as a function of both scale and redshift 
\citep[e.g.][]{2005MNRAS.362..505C, 2003MNRAS.345..923V}.
Previous studies  have calibrated semi-analytic models for 
the halo mass bias from simulations using either a Friends-of-Friends (FOF) halo finding algorithm \citep{1998ApJ...503L...9J,Sheth:1999su,2004MNRAS.355..129S,2005ApJ...631...41T,2010MNRAS.402..191P} or a spherical overdensity (SO) halo finder
\citep{2003ApJ...584..702H,2010MNRAS.402..589M,2010ApJ...724..878T}.
In the FOF approach, particles are simply linked together in a percolation scheme which tracks iso-density contours. The main advantage of this method
is that it makes 
no assumptions about halo geometry and tracks the shapes of bound objects faithfully.
In the SO approach halos are identified as isolated spheres around density peaks where the mass of a halo 
is defined by the overdensity relative to the background. 
Simulations have shown that the mass function and bias for FOF and SO defined halos
can differ substantially at the high mass end where the FOF algorithm tends to spuriously group distinct halos together \citep{2009ApJ...692..217L,2010ApJ...724..878T}.
Recent analytical advancements to the excursion set theory of halo formation which 
accounts for both non-Markovian walks and 
stochastic barriers have been developed \citep{2010ApJ...711..907M} but 
have yet to be rigously tested or calibrated against simulations.
In this paper we re-visit some of these models and compare their 
predictions with the measured bias for FOF halos in the {\sc mxxl } 
simulation at $z=0$ and $z=0.7$ for a much wider range of halo masses 
than previously explored at both redshifts (e.g. \citealt{Angulo:2007fw}).

\citet{2011ApJ...726....5O} carried out a detailed analysis of the redshift space clustering of 
dark matter halos and the systematic effects on measuring the growth rate parameter taking into account 
uncertainties in the halo mass bias. Recent advancements in modelling redshift space distortions, where the 
apparent positions of galaxies are altered along the line of sight by their intrinsic velocities \citep{Kaiser:1987qv},
have shown that taking into account  nonlinearities in the velocity field provides an improved model for the power spectrum on quasi-linear scales \citep{Scoccimarro:2004tg, 2011MNRAS.410.2081J}.
These studies focused on the redshift space distortion effects in the dark matter only and assume that halo velocities trace the dark matter velocity field faithfully. 

Here we present, for the first time, the halo velocity divergence 
power spectra for different halo mass bins and show that there 
is a significant sampling bias compared to the  dark matter velocity 
power spectrum. Measuring the velocity field from simulations
has been shown to be extremely sensitive to resolution effects \citep{2009PhRvD..80d3504P,2011MNRAS.410.2081J}.
The high force and mass resolution in the {\sc mxxl} simulation allows us to 
accurately probe the extent of this velocity bias for different halo masses 
as a function of scale, redshift and number density. 
This has not been possible before for such a broad range of halo masses. 
Accounting for and modelling this bias in improved redshift space 
distortion models is left to future work.

The attainable precision of cosmological parameters extracted from 
clustering statistics is also limited by the galaxy shot noise which 
is often modelled using Poisson statistics. Following the work of 
\citet{2009PhRvL.103i1303S} we investigate if a mass dependent weighting of the density field can be used to suppress the shot noise in the clustering signal of high mass halos compared to the Poisson signal. This method relies on the
assumption that on large scales the halo or galaxy cross correlation 
coefficient is unity assuming a deterministic  relationship between the 
dark matter and halo density fields.

The {\sc mxxl} simulation \citep{2012MNRAS.426.2046A} is one of the largest high-resolution cosmological simulations to date, employing over 300 billion particles to model the evolution of the matter distribution in a volume of almost $70 {\rm Gpc}^{3}$. The 
{\sc mxxl} run complements previous simulations of the same cosmology in different 
box sizes with different particle numbers, the Millennium and Millennium-II simulations 
\citep{Springel:2005nw,2009MNRAS.398.1150B}. 
At present the largest simulations carried out such as the MICE Grand Challenge \citep{2013arXiv1312.1707F} of 70 billion dark-matter particles in a (3 $h^{-1}$Gpc$)^3$ 
comoving volume;
the Dark Energy Universe Simulation Full Universe Run \citep{2012arXiv1206.2838A} of 550 billion particles in a (21 $h^{-1}$Gpc$)^3$ comoving volume;
the MultiDark simulation \citep{2012MNRAS.423.3018P} of 3840$^3$ particles in a (2.5 $h^{-1}$Gpc$)^3$ comoving volume;
the DarkSky simulation \citep{2014arXiv1407.2600S} of $\sim 10240^3$ particles in a volume $8h^{-1}\mathrm{Gpc}$ on a side
 or the Horizon Run 3 simulation
\citep{2011JKAS...44..217K} of 375 billion particles in (10.8 $h^{-1}$Gpc$)^3$ comoving volume, cannot match the {\sc mxxl} simulation in both mass and force resolution, 
which allows us to accurately model halo masses
and velocities from $10^{12} - 10^{15} h^{-1}M_{\sun}$ over a range of redshifts.
\begin{table}
\caption{Mass bins and number densities for the halo samples shown in Fig. \ref{fig:pk} for $z=0$ and $z=0.7$. 
Note the last bin of masses in the range $3-6 \times 10^{14}$$( h^{-1}M_{\sun})$ is not plotted at $z=0.7$.
}
\begin{center}
\begin{tabular}{@{}lrl}
\hline
mass range\phantom{00000}    &  \phantom{0000}number  density  &  \hspace{-0.3cm}$(h/$Mpc$)^3$ \\
\end{tabular}
\begin{tabular}{@{}lcc}

$( h^{-1}M_{\sun})$    &  $z = 0$ & $z=0.7$ \\
\hline
$> 1 \times 10^{12}$  &    $3.54 \times 10^{-3}$  &   - \\
$1-3 \times 10^{12}$  &   $2.26  \times 10^{-3}$ & $2.27 \times 10^{-3}$ \\
$7-9 \times 10^{12}$  &   $1.22 \times 10^{-4}$ & $1.07 \times 10^{-4}$\\
$1-3 \times 10^{13}$  &    $2.47 \times 10^{-4}$ &$1.96 \times 10^{-4}$\\
$ 5-7 \times 10^{13}$    & $2.38 \times 10^{-5}$ & $1.42 \times 10^{-5}$\\
$9\times 10^{13} - 3 \times 10^{14}$ & $2.82 \times 10^{-5}$ & $1.19 \times 10^{-5}$ \\
$3-6 \times 10^{14}$   &  $3.92 \times 10^{-6}$ &  - \\
\hline
\end{tabular}
\end{center} 
\label{table:masses}
\end{table}

\begin{figure*}
\center
{\epsfxsize=18.5truecm
\epsfbox[53 367 510 523]{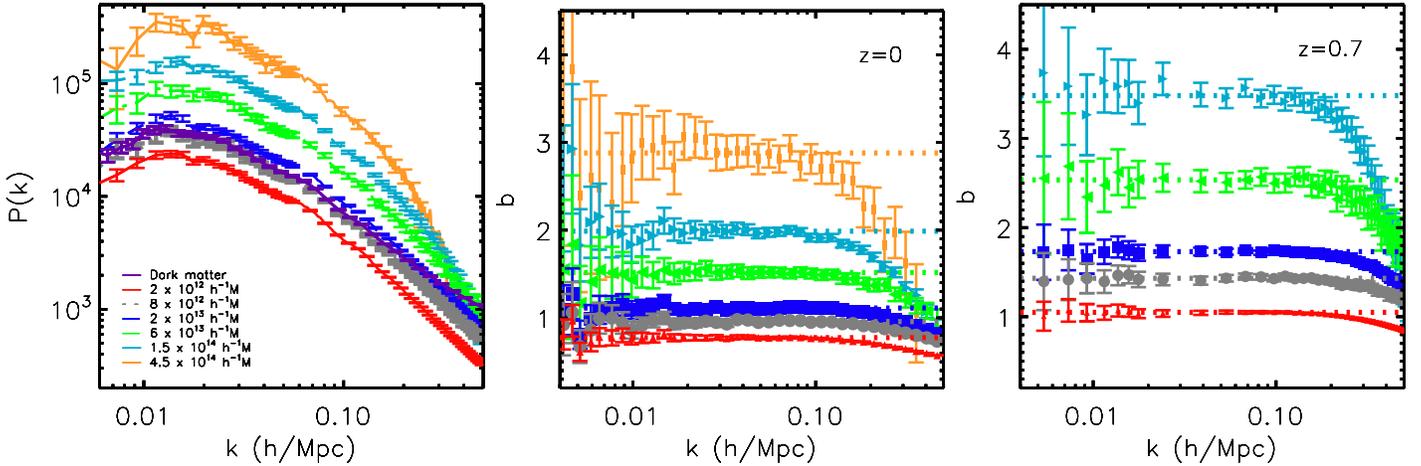}}
\caption{ Left: The halo power spectra for different halo mass ranges measured from the {\sc mxxl} simulation at z=0.
Note that not all error bars are plotted after $k = 0.06 (0.02) h$Mpc$^{-1}$ in the left (right) panel for clarity.  Middle: The $z=0$ measured halo bias for different mass ranges where
$b = \sqrt{P_{\rm hh}/P_{m}}$. The horizontal lines in each case represent the best fit value for the bias over the range
$k < 0.1 h$Mpc$^{-1}$.
Right: The halo bias measured at $z=0.7$. The colour coding for each mass bin is given by the legend in the left panel.
\label{fig:pk}}
\end{figure*}

This paper is organised as follows: In Section~\ref{section:mxxl} 
we describe the {\sc mxxl} $N$-body simulation used in this paper. 
In Section~\ref{section:bias} we analyse the halo matter power spectra 
for different mass bins at redshift $z=0$ and $z=0.7$. These redshifts 
are chosen to be revelvant to current and future redshift surveys. 
We compare the measured linear bias from the {\sc mxxl} simulation to 
different models for the bias at both redshifts.
In Section~\ref{section:shot} we 
examine whether the shot noise for a high mass sample of halos can be reduced using a mass weighting method compared with Poisson shot noise estimates.
In Section~\ref{section:velocity}
we present
the measured velocity divergence power spectra for the different mass bins measured from the {\sc mxxl } simulation at  redshift $z=0$ and $z=0.7$
and compare with theoretical models for the dark matter velocity field.
Our conclusions and summary are presented in Section~\ref{section:summary}.

\section{The {\sc mxxl} N-body simulation }\label{section:mxxl}

The {\it Millennium-XXL} ({\sc mxxl}) simulation  follows the evolution of the matter distribution within a cubic region of $4.11$Gpc (3$h^{-1}$)Mpc on a side using $6720^3$ particles  \citep[see ][for full details]{2012MNRAS.426.2046A}.
The simulation volume is equivalent to that of the full sky out to redshift $0.7$. The {\sc mxxl} run particle mass is $m_p = 8.456 \times 10^9 h^{-1}M_{\sun}$.
The {\sc mxxl} adopts the same $\Lambda$CDM cosmology as in \citet{Springel:2005nw,2009MNRAS.398.1150B} which faciliates the use of all three simulations for comparative studies on galaxy formation in simulations. The cosmological parameters of the simulation are $\Omega_m = 0.25, \Omega_b = 0.045, \Omega_{\Lambda}=0.75, \sigma_8 = 0.9$ and
$H_0 = 73$km s$^{-1}$Mpc$^{-1}$. Although the power spectrum normalization $\sigma_8$ is somewhat high compared to current estimates \citep{2010arXiv1001.4538K} the
 theoretical models for the halo bias and velocity statistics considered in this work have previously been tested using simulations of varying cosmologies 
\citep{2010MNRAS.401.2181J,2010ApJ...724..878T}. Given the impressive mass and force resolution in the 
{\sc mxxl} simulation, it is interesting to test the validity of these models for halo masses which lie beyond the resolving power of the original simulations used for calibration. 

The initial conditions for the simulation were laid down by periodically replicating a $280^3$ particle cubic glass \citep{White:1996}; \citep[see also][]{Baugh:1995hv}
file twenty-four times in each coordinate direction. The displacement and velocities for each particle at the starting redshift of $z=63$ were then computed using
second-order Lagrangian perturbation theory \citep{1998MNRAS.299.1097S}.
The {\sc mxxl} simulation was run using a  ``lean'' version of the {\sc GADGET}-3 code which is a highly optimised version of the TreePM code {\sc GADGET}-2 
\citep{Springel:2005nw,Springel:2005mi}. The group finder makes use of a Friends-of-Friends (FOF) algorithm \citep{1985ApJ...292..371D}
to locate gravitationally bound structures.

\citet{2012MNRAS.426.2046A} measured the mass function of dark matter 
halos using the $z=0$ output of the {\sc mxxl} over the mass 
interval 
$2 \times 10^{11} h^{-1} M_{\odot} - 3 \times 10^{15} h^{-1} M_{\odot}$, 
and combined this with the other simulations in the Millennium suite to 
calibrate a new fitting formula to describe the mass function. 
Angulo et~al. also demonstrated that the matter power spectrum 
could be measured from the {\sc mxxl} over the wavenumber 
range $ 2 \times 10^{-3} h {\rm Mpc}^{-1}$ to $ 10 h {\rm Mpc}^{-1}$, 
demonstrating the huge dynamic range of the simulation. 
The {\sc mxxl} has also been used with semi-analytical galaxy formation 
models to study the appearence of the baryonic acoustic oscillation peak 
when using different galaxy tracers \citep{2014MNRAS.442.2131A}.

\section{The spatial distribution of dark matter halos }\label{section:bias}

\begin{figure*}
\center
{\epsfxsize=17.5truecm
\epsfbox[75 368 509 634]{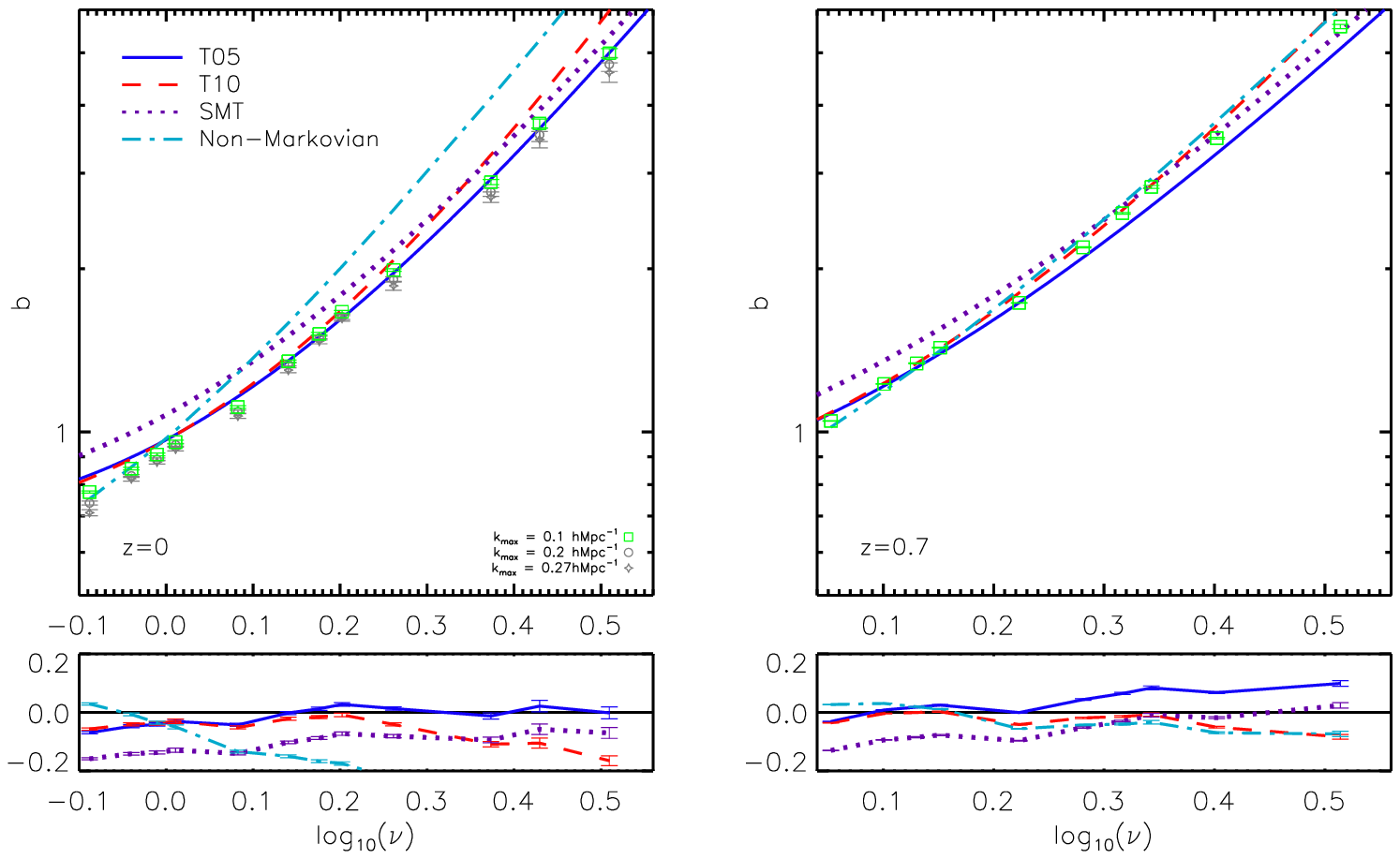}}
\caption{ Left: The linear halo mass bias $b({\rm log}\nu)$ measured from 
the {\sc mxxl} simulation at $z=0$ where $b = \sqrt{P_{\rm hh}/P_{m}}$ 
using $0.004<k (hMpc^{-1}) < 0.1$ is shown as green squares. 
The lines show analytic models from \citet{2005ApJ...631...41T} (blue solid), 
\citet{2010ApJ...724..878T} (red dashed), \citet{Sheth:1999su} (purple dotted) 
and the model from \citet{2011MNRAS.411.2644M} (cyan dot-dashed) which includes 
non-Markovian terms and a stocastic barrier.
Right: The halo bias at $z=0.7$. All model predictions were generated by scaling 
the variance of the linear density field by the ratio of the linear growth at $z=0.7$ and $z=0$.
The \citet{2011MNRAS.411.2644M}  model has been plotted at this redshift (cyan dot dashed lines) using the best fit values to the {\sc mxxl} $b-{\rm log}\nu$ relation.
In both cases the lower panels show the ratio of the model predictions to the mass bias measured
from the simulation.
\label{fig:bias}}
\end{figure*}

In this section we present the measured power spectra of various halo mass samples from the {\sc mxxl} simulation at $z=0$ and $z=0.7$. We focus first on comparing the halo power spectrum with that of the matter distribution, 
as quantified through the halo mass bias (Section 3.1). We then test a prescription for 
suppressing the shot noise in the power spectrum of a halo sample which is a modification of the method 
proposed by \citet{2009PhRvL.103i1303S} (Section 3.2).

\subsection{Bias of dark matter halos }

We analyse the linear bias measured from the ratio of the halo auto and mass power spectra $b \equiv  (P_{\rm hh}/P_{\rm m})^{1/2}$ as a function of scale and compare the predictions
for the bias - peak height relation with commonly used models.
The power spectrum was computed by assigning the particles to a mesh using the cloud in cell (CIC) assignment scheme \citep{1981csup.book.....H} 
and then performing a fast Fourier transform on the density field. To restore the resolution 
of the true density field this assignment scheme is corrected for by performing an 
approximate de-convolution \citep{1991ApJ...375...25B}.
Throughout this paper the fractional error on the power spectrum plotted is given by 
$\sigma_P/P = (2/N)^{1/2}(1 + \sigma_n^2/P)$ where $N$ is the number of modes measured in a spherical shell of width $\delta k$  and $\sigma_n$ is the shot noise \citep{1994ApJ...426...23F}. This number depends upon the survey volume, $V$, as
$N = V 4\pi k^2 \delta k/(2\pi)^3$.

In Fig. \ref{fig:pk} we show the $z=0$ halo power spectra for the halo samples listed in Table \ref{table:masses}. The dark matter power spectrum is shown as a purple solid line in this figure. 
In the middle panel we show the halo bias at $z=0$ for each halo sample evaluated as $b = \sqrt{P_{\rm hh}/P_{m}}$.
The best fitting value for the bias over the range $0.004<k (h$Mpc$^{-1}) < 0.1$ is shown as horizontal lines for each sample.
These ratios are remarkably flat over the range
 $k < 0.2 h$Mpc$^{-1}$ for
masses  $< 2 \times 10^{13}h^{-1}M_{\sun}$ and $k < 0.1 h$Mpc$^{-1}$ for masses $> 6\times 10^{13}h^{-1}M_{\sun}$ at $z=0$ 
in agreement with the work of \citet{2011ApJ...726....5O}.
At a higher redshift of $z=0.7$ this bias is  scale independent for all masses at $k < 0.1 h$Mpc$^{-1}$ although the scale dependence is more pronounced on quasi-linear scales compared to redshift zero.

In Fig. \ref{fig:bias} we show the linear halo mass bias $b$, as a function of $\log \,\nu$ where $\nu = \delta_c/\sigma(R)$,
 measured from the {\sc mxxl} simulation at $z=0$ as green squares. Here $\sigma(R)$ is the variance of the smoothed density field defined as
\begin{eqnarray}
\sigma^2(R) = \frac{1}{\left(2\pi^2\right)}\int_0^{\infty} {\rm d ln}k k^2 P(k) W^2(k,R) 
\end{eqnarray}
where $W(k,R)$ is the Fourier transform of a top hat window function and $\delta_{c}$ is the threshold for perturbation collapse in linear theory. 
The best fitting value for the bias was obtained using the range $0.004<k (h$Mpc$^{-1}) < 0.1$. We find that the estimated bias is sensitive to the maximum wavenumber used in the fit; extending this  to smaller scales where non-linear bias is present decreases the bias as shown for 
 $k_{\tiny \rm max} = 0.2 h$Mpc$^{-1}$ (grey circles) and $k_{\tiny \rm max} = 0.27 h$Mpc$^{-1}$ (grey stars). 
When fitting a linear scale independent bias to the simulation results we find a gradual decline in the best fit value with increasing $k_{\tiny \rm max}$. This indicates that the bias becomes scale dependent. Unfortunately there is not a sudden jump in the recovered bias which would indicate a good point at which to limit the range of $k$-values used in the fit.  
Fig.~\ref{fig:bias} compares various analytic models for the halo 
mass bias to the {\sc mxxl} measurements.  
The \citet{Sheth:1999su} model improves on the halo bias predictions 
assuming spherical collapse by using a moving barrier whose scale-dependent 
shape is motivated by the ellipsoidal gravitational collapse model. 
It is well know that this model overpredicts the bias at 
the low mass end while overall it matches the results of 
simulations within 20\% in agreement with our results shown in Fig. \ref{fig:bias} \citep{2005ApJ...631...41T,2004MNRAS.355..129S,2010MNRAS.402..191P}.
The models of \citet{2005ApJ...631...41T} and \citet{2010ApJ...724..878T} 
represent updated fitting formulae calibrated using halos defined using a friends-of-friends (FOF) and spherical overdensity (SO) algorithm 
respectively. 
The SO algorithm identifies halos as isolated density peaks, whose masses are determined by the 
overdensity $\Delta$, defined here as the mean interior density relative to the background density.
We have used $\Delta = 200$ times the critical density in the \citet{2010ApJ...724..878T} model 
as this overdensity is close to the overdensity of halos identified with the 
FOF algorithm with a linking length of 0.2 \citep{1985ApJ...292..371D}.
The discrepancy between the \citet{2010ApJ...724..878T} model predictions and the {\sc mxxl } bias relation at the high mass end is most likely due to
difference in the halo finder used in each case. 
As shown in \citet{2008ApJ...688..709T} and \citet{2009ApJ...692..217L}, a SO finder would identify a significant fraction of FOF halos as two distinct density peaks.
This artifact of FOF linking increases the 
abundance of massive FOF halos relative to the abundance of SO halos and reduces the bias as seen in Fig.~\ref{fig:bias}
 \citep[see also figure 3 in][]{2010ApJ...724..878T}.
Overall we find good agreement to within $10$\% between the {\sc mxxl} bias relation and the model from \citet{2005ApJ...631...41T}.
The low mass end of the bias relation is well fit to within a few percent 
by the two parameter model of \citet{2011MNRAS.411.2644M}  which a incorporates non-Markovian extension of 
the excursion set theory with a stochastic barrier \citep[see also][]{2010ApJ...711..907M}. In Fig. \ref{fig:bias} we use $\kappa = 0.23, a=0.818$ where
 the two parameters $\kappa$ and $a$ describe 
 the degree of non-Markovianity and the degree of stochasticity of the barrier respectively \citep[see][for more details]{2011MNRAS.411.2644M}.

The right hand side of Fig.~\ref{fig:bias} compares the bias measured from the {\sc mxxl} 
simulation at $z=0.7$. In this case the \citet{2005ApJ...631...41T} provides a reasonable 
match to the bias of halos corresponding to modest peak heights. For rarer peaks, 
the \citet{Sheth:1999su} works better at this redshift.

\begin{figure*}
\center
{\epsfxsize=17.5truecm
\epsfbox[77 364 555 579]{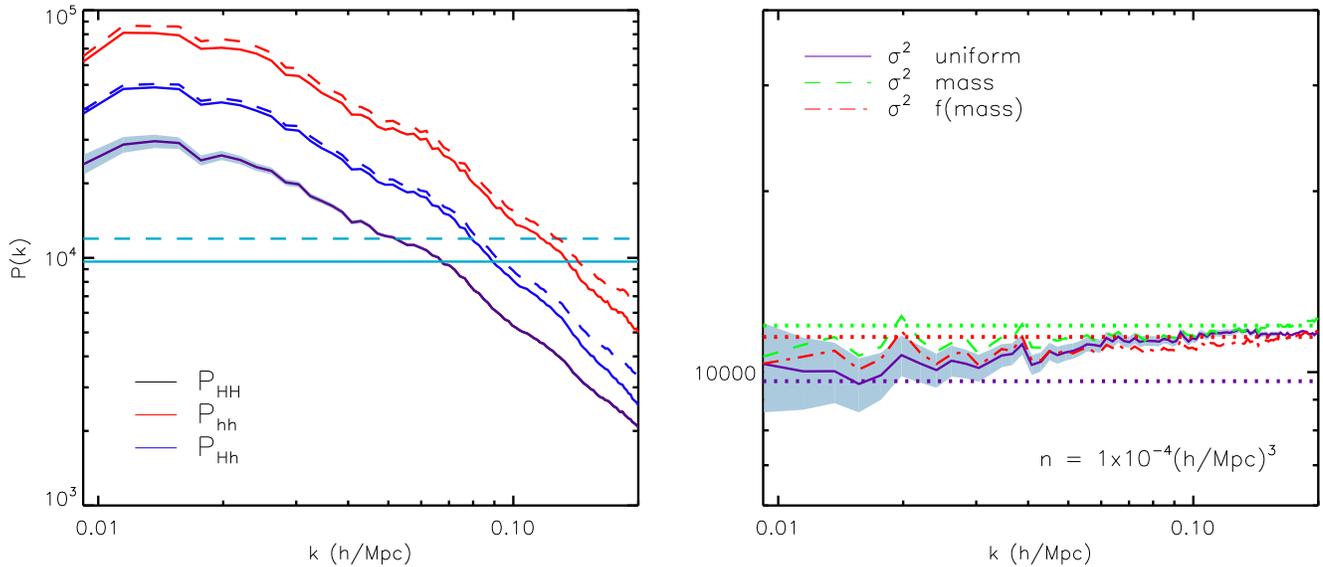}}
\caption{Left: The measured power spectrum with uniform weighting for all halos  with  masses $M>10^{12}M_{\sun}/h$ and halos with $M \in 3\times10^{13} - 1\times10^{14}M_{\sun}/h$
are shown as a solid purple and red line respectively. The cross spectrum with uniform weighting for these two tracers is shown as a blue solid line.
Power spectra using mass weightings for the $M \in 3\times10^{13} - 1\times10^{14}M_{\sun}/h$ sample are shown as dashed lines. The expected Poisson shot noise for the uniform and
mass weighting schemes are shown as horizontal solid and dashed lines. The grey shaded regions show errors on the uniform weighted $P(k)$ for the halo sample $M>10^{12}M_{\sun}/h$.
Right: The measured noise and expected Poisson shot noise for different weighting schemes are shown as solid and dotted lines respectively.
\label{fig:shotnoise}}
\end{figure*}

\subsection{Minimising shot noise }\label{section:shot}

The two main sources of error in a measurement of the power spectrum are cosmic variance, 
due to a finite number of modes available on large scales with which to determine the variance of 
the field, and the shot noise due to the discrete sampling of the density field using 
galaxies or halos. If we assume Poisson statistics then the shot noise error equals 
the inverse of the number density which can be simply substracted from the overall measurement.
Within the halo model where all dark matter lies in collapsed halos of different masses 
there should be significant halo exclusion effects for the most massive halos which will 
cause discrete sampling effects to deviate from Poisson statistics.
\citet{2009PhRvL.103i1303S} proposed accounting for this difference using mass weighting 
schemes to boost the clustering signal of a halo sample resulting in a shot noise term 
which is lower then predicted from $1/\bar{n}$ Poisson statistics. Here we make use of 
the cross correlation power spectra between a high number density halo sample, whose 
shot noise is negligible, and a high-mass, low number density sample 
with $\bar{n} \sim 10^{-4} (h/$Mpc$)^3$.

Consider the cross correlation between the dark matter and a tracer, which has 
overdensity $\delta_h$ and noise $n$, where the cross correlation coefficient is
\begin{eqnarray}
r \equiv \frac{P_{hm}}{\sqrt{P_{hh}P_{m}}}.
\end{eqnarray}
Here $P_{hh}, P_{m}$ are the auto power spectra for the halos and mass and 
$P_{hm} = \langle \delta_h\delta_m\rangle$ is the cross power spectrum. 
Given $r = 1$ we can re-write this in terms of the shot noise 
$\sigma^2 = \langle n^2 \rangle$ where $P_{hh} = \langle \delta_h^2\rangle - \sigma^2$,
\begin{eqnarray}
\sigma^2 = \langle \delta_h^2\rangle -  \frac{P^2_{hm}}{P_{m}}.
\end{eqnarray}
Everything on the RHS of the above equation can be measured from simulations 
(or from surveys by combining clustering and lensing measurements) and the resulting $\sigma^2$ can be compared with the Poisson prediction as a function of scale.
In the case of uniform weighting for each halo in the sample the Poisson prediction is $\sigma^2_{\tiny expected} = 1/{\bar{n}}$. If we weight each halo by its mass, using weights 
$w_i$ then the expected shot noise is $\sigma^2_{\tiny expected} = V \sum_i w_i^2/(\sum_i w_i)^2$.

Here we modify this approach as follows.
Using two halo samples labelled \\
$H : $ all halos with $>10^{12}h^{-1}M_{\sun}$ \\
$h : $ halos with mass $ \in 3\times10^{13} - 1\times10^{14}h^{-1}M_{\sun}$\\
 we can define a cross correlation coefficient between them as
\begin{eqnarray}
r_{Hh} \equiv \frac{P_{Hh}}{\sqrt{P_{hh}P_{HH}}}
\end{eqnarray}
where $P_{hh} = \langle \delta_h^2\rangle - \sigma_h^2$ and $P_{HH} = \langle \delta_H^2\rangle - \sigma_H^2$ and $P_{Hh} = \langle \delta_h \delta_H \rangle$ is the cross spectrum and we have assumed that the noise for each tracer is uncorrelated with the other i.e. $\sigma^2_{Hh} = \langle n_H n_h\rangle = 0$.
We make two assumptions: firstly as the halo sample $H$ is large we assume that the noise term $\sigma^2_{HH}$ in the above equation is small and neglect it, secondly we assume that 
on large scales there is a deterministic relationship between these two tracers such that 
the cross correlation coefficient is equal to one $r_{Hh} = 1$ \citep[see e.g.][]{2008MNRAS.385.1635S}.
We can then write the shot noise term for the $h$ halo sample as
\begin{eqnarray}
\sigma^2_{hh} = \langle \delta_h^2\rangle -  \frac{P^2_{Hh}}{P_{HH}}.
\end{eqnarray}
Using these two halo samples from {\sc mxxl} we can compare the measured shot noise from the above equation and compare it with the Poisson prediction in the case of uniform or mass weighting schemes.

\begin{figure*}
\center
{\epsfxsize=17.5truecm
\epsfbox[71 370 510 686]{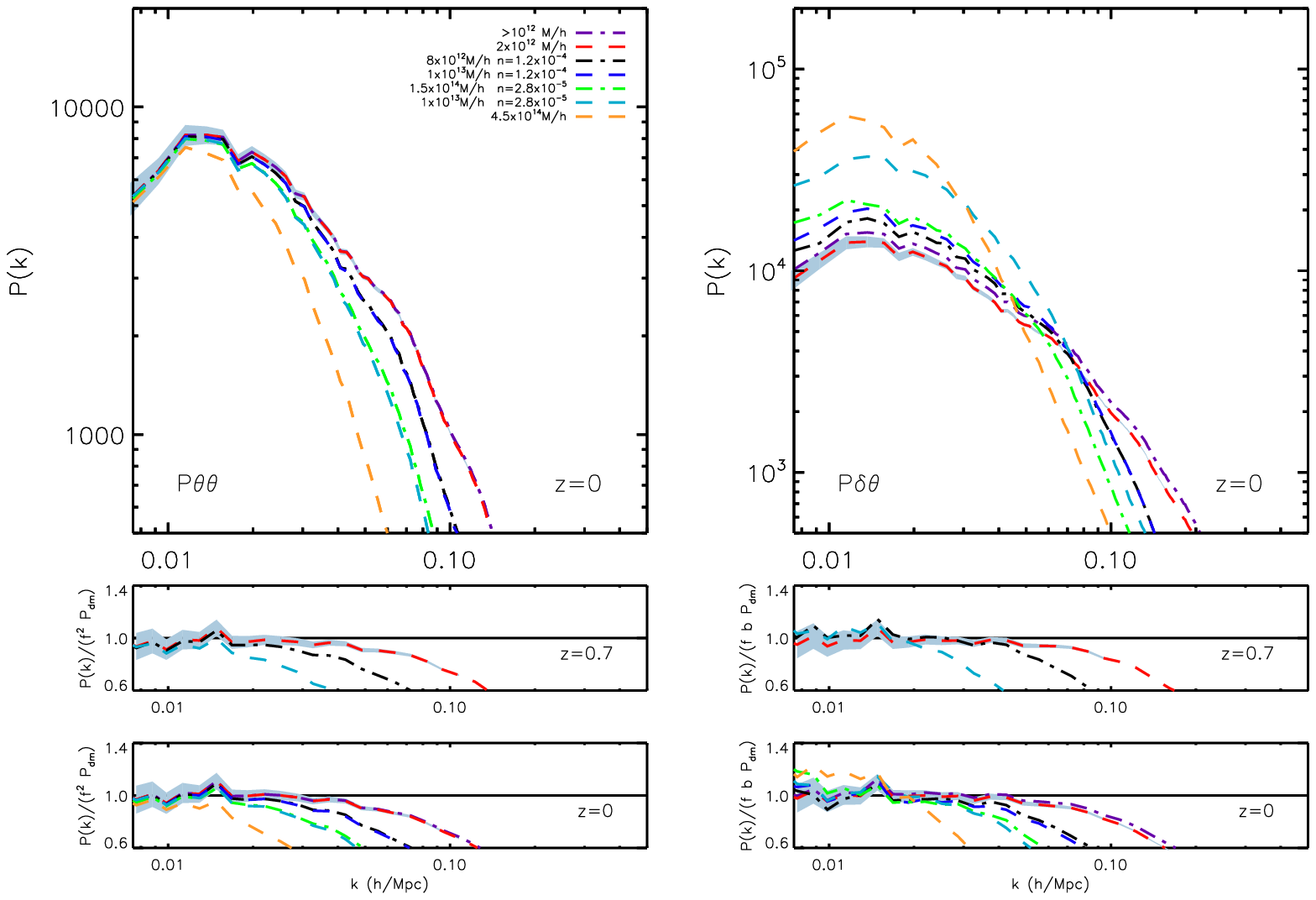}}
\caption{ Upper panels: The auto (left) and cross (right) velocity divergence power spectra measured from the {\sc mxxl} simulation at $z=0$
for different mass bins and number densities as shown in the legend.
For clarity only error bars  for the 2$\times 10^{12}h^{-1}M_{\sun}$ bin
are plotted (grey shaded region).
Lower panels: The $z=0$ velocity divergence power
spectra normalised by $f^2 P_{\rm \tiny dark matter}$ and $f b P_{\rm \tiny dark matter}$ for the auto and cross power spectra respectively.
Note the dark blue and grey dashed lines correspond to different mass ranges but equal number densities.
Middle panels: Same ratios as the lower panels at $z=0.7$.
\label{fig:vel}}
\end{figure*}

\begin{figure*}
\center
{\epsfxsize=17.5truecm
\epsfbox[71 370 510 686]{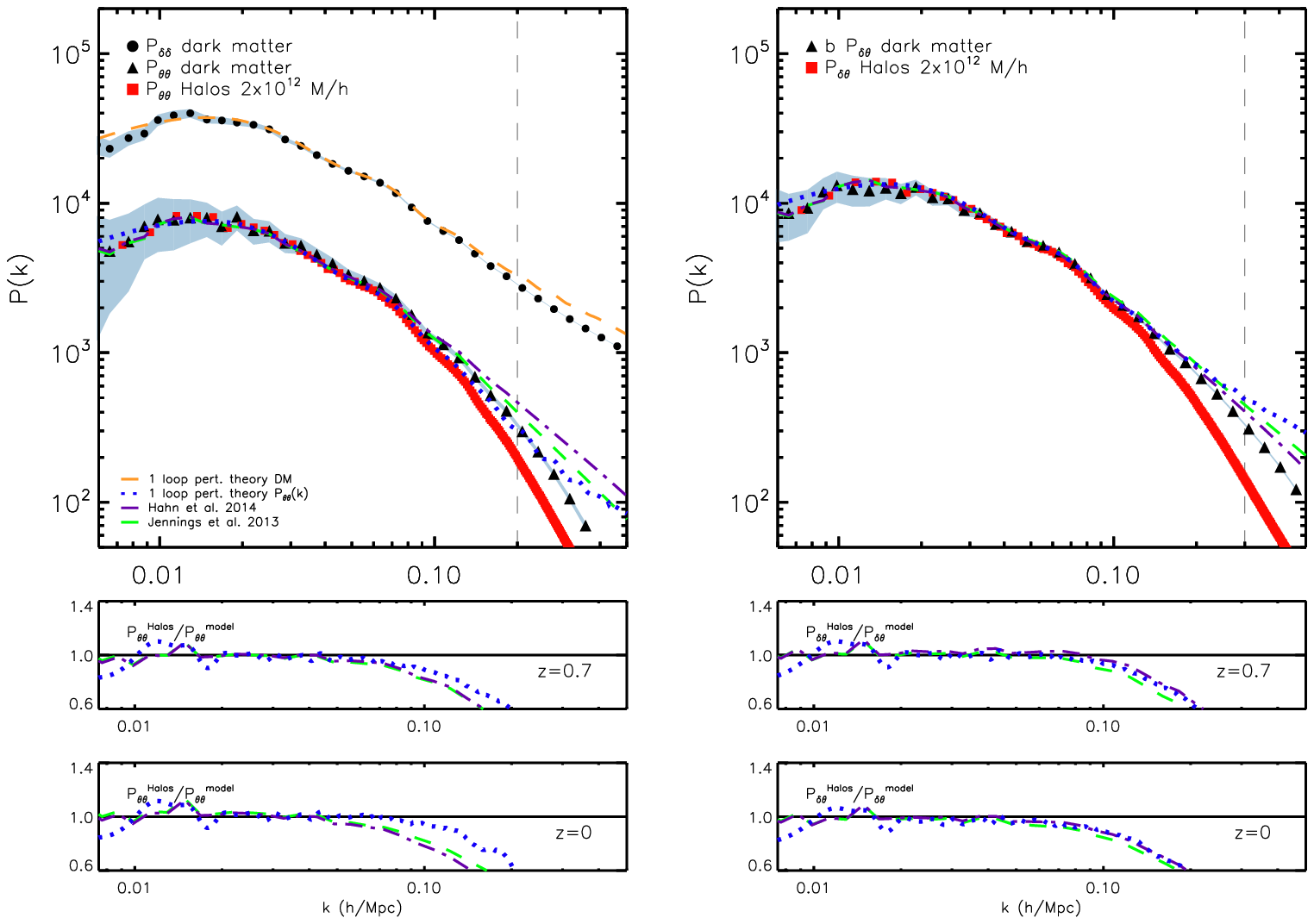}}
\caption{ Upper panels: The {\sc mxxl} matter (black circles) and velocity (black triangles)
power spectra for the dark matter at $z=0$. The velocity divergence power spectra for the 2$\times 10^{12}h^{-1}M_{\sun}$ mass bin are plotted
as red squares in both panels. The error bars for the dark matter power spectra are plotted as grey shaded regions.
The velocity divergence power spectrum predictions from 1 loop perturbation theory, Jennings et al. 2013 and Hahn et al. 2014 are shown as blue dotted, green dashed and purple dot dashed lines respectively in all panels. 
The dark matter $P(k)$ predicted from perturbation theory is shown as an orange dashed line in the upper left panel only.
 Lower panels: The ratio between the power spectra measured for the 2$\times 10^{12}h^{-1}M_{\sun}$ mass bin halo sample (see labels) and the different model predictions
at $z=0$. Middle panels: same ratios as shown in the lower panels but for  $z=0.7$.
\label{fig:velocity_models}}
\end{figure*}

In Fig. \ref{fig:shotnoise} the measured power spectrum with uniform 
weighting for all halos  with  masses $M>10^{12}h^{-1}M_{\sun}$ and halos with $M = 3\times10^{13} - 1\times10^{14}h^{-1}M_{\sun}$
are shown as a solid purple and red line respectively. The cross spectrum with uniform weighting for these two tracers is shown as a blue solid line.
Power spectra using mass weightings for the $M \in 3\times10^{13} - 1\times10^{14}h^{-1}M_{\sun}$ sample are shown as dashed lines. The expected Poisson shot noise for the uniform and
mass weighting schemes are shown as horizontal solid and dashed lines.
As can be seen from this plot, the mass weighting scheme boosts the Poisson shot noise term but also boosts the clustering signal.

In the right panel the measured noise and expected Poisson shot noise for different weighting schemes are shown as solid and dotted lines respectively.
Here the $f(\rm mass) $ weighting scheme is the one suggested in \citet{2009PhRvL.103i1303S} where $f(M) = M/(1 + \sqrt(M/10^{14}h^{-1}M_{\sun}))$
As can be seen from this figure on large scales the 
measured and expected shot noise in the case of uniform weighted 
agree on large scales (purple solid and dotted lines) but this agreement breaks down as we 
go to smaller non-linear scales. This may be due to stocasticity on small scales as $r$ differs from unity or the fact that assuming
Poisson shot noise overestimates the noise levels for highly biased tracers as found in \citet{2009PhRvL.103i1303S}.
Using either the mass or $f(M)$ weighting schemes we find a small (factor of 1.5) 
reduction in the measured shot noise compared to the expected value from Poisson statistics. These improvements are
small compared to the factor of three reduction in shot noise which \citet{2009PhRvL.103i1303S} found when using the cross correlation between a halo sample and the dark matter field.
Although this approach does not yield such a large reduction in shot noise the main advantage of this
method is that the dark
matter density field does not need to be estimated in contrast to the method presented in \citet{2009PhRvL.103i1303S}.

\section{Velocity bias }\label{section:velocity}

In this section we examine the statistics of the velocity field measured from the dark matter and halo populations in the {\sc mxxl} simulation through 
the auto, $P_{\theta \theta}$, and cross, $P_{\delta \theta}$, power
spectra for the velocity divergence $\theta \equiv \vec{\nabla} \cdot \vec{v}/(aH)$, where $a$ is the scale factor and $H$ is the Hubble rate.
These two power spectra are important in many models for redshift space distortions 
\citep[see e.g.][]{Scoccimarro:2004tg,2009MNRAS.393..297P,2011MNRAS.410.2081J,2011MNRAS.416.2291T,2012MNRAS.427..327D}.
Any bias between the velocity divergence power spectra for a galaxy/halo population
and the underlying dark matter would have important implications for cosmological parameters extracted assuming that a tracer population follows the dark matter exactly.
To our knowledge this is the first time that these power spectra have been analysed for different halo populations using simulations.  

Measuring the velocity power spectrum accurately from N-body simulations can be difficult
as both mass and volume weighted approaches can involve significant noise and biases on small 
scales \citep{2011arXiv1105.0370C,2003A&A...403..389P,Scoccimarro:2004tg,2009PhRvD..80d3504P,2011MNRAS.410.2081J,2012MNRAS.427L..25J,2014arXiv1404.2280H}. The method suggested by \citet{Scoccimarro:2004tg}
allows a mass weighted velocity field to be constructed but is limited by the fact that it is the momentum field
which is calculated on a grid and so the velocity field in empty cells is artificially set to zero \citep{2009PhRvD..80d3504P}.
Another limitation of this method is that most calculations require the volume weighted velocity field instead of the mass weighted field.
Using a Delaunay tessellation
of a discrete set of points allows the desired volume weighted velocity field to be constructed accurately on small scales. 
We use the publicly available {\sc dtfe} code \citep{2011arXiv1105.0370C}
to construct the velocity divergence field for our halo samples directly.
This code  constructs the Delaunay tessellation from a discrete set of points and interpolates the field values onto a user defined grid.
The density field is interpolated onto the
grid using the cloud-in-cell assignment scheme. The resolution of the mesh means that
mass assignment effects are negligible on the scales of interest here.

Given the large number density of particles in the {\sc mxxl} simulation it is numerically infeasible to run the 
{\sc dtfe} code on the dark matter. Instead, we adopt the  mass weighted method suggested by \citet{Scoccimarro:2004tg} to measure 
$P_{\theta \theta}$ and $P_{\delta \theta}$ for the dark matter particles both at $z=0$ and $z=0.7$ using a 1024$^3$ grid.
Using smaller volume $\Lambda$CDM simulations in a box of 1500$h^{-1}$Mpc on a side and 1024$^3$ particles 
from \citet{2012MNRAS.427L..25J}, we have verified that this mass weighted method
agrees with the {\sc dtfe} dark matter velocity field up to $k\sim0.2h/$Mpc for $P_{\theta \theta}$ and
$k\sim0.3h/$Mpc for $P_{\delta \theta}$. We will restrict our comparison between the velocity statistics for dark matter and halos to this range where
the mass weighted method has converged.
To account for both aliasing and shot noise effects 
on both the halo and dark matter velocity power spectra we have verified that increasing the size of the grid used ($1024^3$) has no effect on the measured power
over the range of scales we consider in this work.

Fig \ref{fig:vel} shows the halo velocity power spectra $P_{\theta \theta}$ (left panel) and
$P_{\delta \theta}$ (right panel) at $z=0$ for different mass ranges and number densities 
given in the legend. For clarity we only plot the error bars for the 2$\times 10^{12}h^{-1}M_{\sun}$ bin as a grey shaded region.
There is a clear difference in the $P(k)$ measured using different halo samples, 
which increases with increasing mass (decreasing number density) on large scales
$k >0.01 h$Mpc$^{-1}$. As shown in \citet{2009PhRvD..80d3504P}{ and \citet{2011MNRAS.410.2081J} the velocity power spectrum is very senstive to resolution and this trend of increasing bias with an increase in the halo mass is actually due to a decrease in the number density of the velocity field tracers.
We verify that this is indeed a number density bias by matching number densities for different mass ranges and comparing the measured $P_{\theta \theta}$
and $P_{\delta \theta}$. As can be seen from the black dot dashed and blue dashed lines in Fig. \ref{fig:vel}, once
we match the number density for these two different mass bins to $\bar{n} = 1.2 \times 10^{-4}(h/$Mpc$)^3$ we obtain the same 
velocity power spectra. We have also verified this for two mass bins which have different bias factors,
$M  = 1.5 \times 10^{14}h^{-1}M_{\sun}$ ($b\sim 2 $ at $z=0$) and $M  = 1 \times 10^{13}h^{-1}M_{\sun}$ ($b\sim 1 $ at $z=0$),
but the same number density $\bar{n} = 2.8 \times 10^{-5}(h/$Mpc$)^3$ 
(green dashed and dot-dashed lines in Fig. \ref{fig:vel}). Note that the power spectra for all halos with masses
$> 10^{12}h^{-1}M_{\sun}$ (purple dot dashed line) is similar to the sample $1-3 \times 10^{12}h^{-1}M_{\sun}$ (red dashed line) in this figure, which is why their measured velocity $P(k)$ agree.

In the lower four panels in Fig. \ref{fig:vel} we show the ratios 
$P^{\rm \tiny halos}_{\theta \theta}/(f^2 P^{\rm \tiny dark matter}_{\delta \delta})$ and 
$P^{\rm \tiny halos}_{\delta \theta}/(f b P^{\rm \tiny dark matter}_{\delta \delta})$  as a function of scale at $z=0$ and $z=0.7$ where
$f \equiv {\rm d ln}D/{\rm d ln}a$ is the growth rate (logarithmic derivative of the growth factor, $D$) and $b$ is the linear bias for each halo sample
at that redshift. From the panels we can see that the velocity $P(k)$ agree with linear theory predictions 
only on large scales $k = 0.004h$Mpc$^{-1}$ at both redshifts for our halo mass bins 
2$\times 10^{12}h^{-1}M_{\sun}$  ($\bar{n} = 2.26 \times 10^{-3}(h/$Mpc$)^3$)
and 8$\times 10^{12}h^{-1}M_{\sun}$ ($\bar{n} = 1.2 \times 10^{-4}(h/$Mpc$)^3$). Beyond $k = 0.004h$Mpc$^{-1}$ we see a departure from linear 
theory and a difference of $\approx 50$\% between the measured velocity $P(k)$ and linear perturbation theory predictions at $k = 0.1h$Mpc$^{-1}$.
For the 1$\times 10^{13}h^{-1}M_{\sun}$ mass bin the measured $P_{\theta \theta}$
and $P_{\delta \theta}$ only agree with linear theory predictions for $k < 0.002h$Mpc$^{-1}$.
We see the largest deviations for the 4.5$\times 10^{13}h^{-1}M_{\sun}$ mass bin, which we were only able to accurately measure at $z=0$.
It is clear from the ratios in these figures that for small number densities, $\bar{n} \sim 10^{-6}(h/$Mpc$)^3$, the sampling bias is extremely large
and  we do not recover the linear theory prediction for the cross power spectrum on large scales.
Note the agreement between the cross spectra and the preditions of linear perturbation theory for these halo 
velocity divergence power spectra is interesting considering that the linear bias used is defined as an average quantity which 
takes into account stochasticity
 $b \equiv (P_{\rm \tiny halo}/P_{\rm \tiny dark matter})^{1/2}$ 
rather then a local linear variable $b = \delta_{\rm \tiny halo}/\delta_{\tiny \rm dark matter}$ \citep{1999ApJ...525..543M}.

In Fig. \ref{fig:velocity_models} we compare the measured {\sc mxxl} matter (black circles) and velocity (black triangles) 
power spectra for the dark matter
and the 2$\times 10^{12}h^{-1}M_{\sun}$ mass bin velocity $P(k)$ (red squares)  with two  models which have been calibrated from N-body simulations. We also compare
these measured power spectra with the predictions of perturbation theory as in \citet{Scoccimarro:2004tg}.
The vertical dashed line in each panel indicates the maximum wavenumber where our velocity $P(k)$ have converged.
Although the \citet{2012MNRAS.427L..25J} (green dashed line) and the \citet{2014arXiv1404.2280H} (purple dot dashed line)
 formulas where calibrated on simulations of different resolutions
and cosmologies to the {\sc mxxl} simulation, and, furthermore, each study used a different method for determining the velocity field, we find very good agreement between both formula and the measured $P_{\theta \theta}$ and $P_{\delta \theta}$ at $z=0$ and $z=0.7$
for $k < 0.15h$Mpc$^{-1}$.
In agreement with \citet{Scoccimarro:2004tg} we find that 1-loop perturbation theory (blue dotted line)
predictions are accurate for $k<0.1h$Mpc$^{-1}$. On smaller scales perturbation theory over (under) predicts the amplitude of the matter (velocity) power
spectra for the dark matter. 

As shown in Fig. \ref{fig:vel} there is a significant sampling bias between velocity power spectra for mass bins with different number densities.
In Fig. \ref{fig:velocity_models}  it is clear that the dark matter (black circles) and 
the 2$\times 10^{12}h^{-1}M_{\sun}$ mass bin (red squares) velocity power spectra only agree up to $k < 0.08h$Mpc$^{-1}$. Even for the largest number
 density mass bin which we use in this study there is a significant sampling bias between the dark matter and the halo velocity $P(k)$.
In order to highlight the discrepancy between the models, which accurately predict the dark matter $P_{\theta \theta}$ and $P_{\delta \theta}$, and the 
halo velocity divergence power spectra, we plot the ratio of these two power spectra in the lower ($z=0$) and middle ($z=0.7$) panels in Fig. 
\ref{fig:velocity_models}. It is clear from these  ratio plots that all models for the dark matter velocity statistics are biased by approximately 
20\%  for $P_{\theta \theta}$ and approximately 10\% for $P_{\delta \theta}$ 
at $k = 0.1 h$Mpc$^{-1}$  compared to the halo velocity divergence $P(k)$. This discrepancy is significant and will have an impact on cosmological parameter 
inference from e.g. redshift space clustering measurements where  redshift space distortions models assume zero velocity bias.
The question of how to correct for this sampling bias in both power 
spectra as a function of scale is beyond the scope of this work and is left for future study.
Note while writing up this paper we became aware of two recent studies by
\citet{2014arXiv1405.5885B} and  \citet{2014arXiv1405.7125Z} who have also reported 
that there should be a bias in the velocity power spectra. \citet{2014arXiv1405.7125Z}
 report that the velocity divergence auto power spectra
for $\bar{n} \sim 10^{-3}(h/$Mpc$)^3$ tracers should be affected by approximately 10\% at $k = 0.1 h$Mpc$^{-1}$ in agreement with our findings.

\section{Conclusions and Summary }\label{section:summary}

We have measured and tested various models for the linear halo mass bias using measurements of 
 the ratio of the halo auto power spectra from the {\sc mxxl} simulation at redshift 
$z=0$ and $z=0.7$ for different mass bins in the range 2 $\times 10^{12} - 3 \times 10^{15}h^{-1}M_{\sun}$.
In agreement with the work of \citet{Angulo:2007fw} and \citet{2011ApJ...726....5O} 
we find that the assumption of a linear bias is only valid on scales $k < 0.2 h$Mpc$^{-1}$ for 
masses  $< 2 \times 10^{13}h^{-1}M_{\sun}$ and $k < 0.1 h$Mpc$^{-1}$ for masses $> 6\times 10^{13}h^{-1}M_{\sun}$ at $z=0$. At a higher redshift of $z=0.7$ this bias is remarkably scale independent for all masses at $k < 0.1 h$Mpc$^{-1}$ although the scale dependence is more pronounced on quasi-linear scales compared to redshift zero. 
When fitting for a linear scale independent bias we find a gradual decline in the best fit value with increasing $k_{\tiny \rm max}$ instead of a sharp jump which would indicated an obvious scale
dependent bias. 

When plotted as a function of peak height we find that the bias - ${\rm log}\nu$ relation is well fit at $z=0$ by the model of \citet{2005ApJ...631...41T} except for low mass halos 
$<7\times 10^{12}h^{-1}M_{\sun}$ whose bias is overpredicted by the model. We find that the non-Markovian and diffusive barrier model of \citet{2010ApJ...711..907M} is a better fit to the linear bias of these low mass halos.
At redshift $z=0.7$ we find that the linear bias of {\sc mxxl} FOF halos more massive then $10^{13}h^{-1}M_{\sun}$ is better fit by the ellipsoidal collapse model of 
\citet{Sheth:1999su} which is accurate to $\sim 5$\% when fitting over the range $0.004 - 0.1 h$Mpc$^{-1}$. We find that the model of \citet{2010ApJ...724..878T},
 which was calibrated on SO halos,
overestimates the FOF halos from the {\sc mxxl} simulation at both redshifts by approximately 10-20\% over the range of masses we consider.

We have investigated different weighting schemes applied to the dark matter halo power spectra clustering measurements  in order to reduce the shot noise for a high mass
 (low number density) sample. We have modified the approach of \citet{2009PhRvL.103i1303S} who made use of the cross correlation power spectra between the halos and dark matter to measure the actual shot noise (assuming deterministic biasing on large scales). \citet{2009PhRvL.103i1303S} found that mass weighting could lower the shot noise compared with Poisson statistics by a factor of
3 for a $\bar{n} \sim 10^{-4} (h/$Mpc$)^3$ sample. Here we make use of the cross correlation power spectra between a large number density halo sample, whose shot noise is negligible,
and a high mass (low number density) sample with $\bar{n} \sim 10^{-4} (h/$Mpc$)^3$. We find that mass weighting is able to reduce the shot noise of the measured 
power spectra by at most a factor of 1.5 compared to the Poisson estimate. Although this approach does not yield such a large reduction in shot noise the main advantage of this 
method is that the dark 
matter density field does not need to be estimated in contrast to the method presented in \citet{2009PhRvL.103i1303S}.

We have measured the velocity divergence auto, $P_{\theta \theta}$, and cross, $P_{\delta \theta}$, 
power spectra for a range of halo masses from the {\sc mxxl} simulation at redshift $z=0$ and $z=0.7$. This is the first time that these velocity statistics have been presented
and compared with the dark matter velocity power spectra
from a simulation. The high mass and force resolution of the {\sc mxxl} simulation allows us to reconstruct the velocity power spectra for halos masses 
$10^{12} - 6\times 10^{14}h^{-1}M_{\sun}$ up to 
$k = 0.1h$Mpc$^{-1}$ and the dark matter velocity power spectra up to $k = 0.2h$Mpc$^{-1}$ ($k = 0.3h$Mpc$^{-1}$)   for $P_{\theta \theta}$ ($P_{\delta \theta}$) at $z=0$.
We find that there is a significant sampling bias in both velocity divergence power spectra at $z=0$ and $z=0.7$ which decreases the measured power compared to
 the dark matter velocity $P(k)$
by approximately 20\% at $k=0.1h$Mpc$^{-1}$ for a $\bar{n} = 2 \times 10^{-3} (h/$Mpc$)^3$ sample. This sampling bias increases to $\sim 40$\% for a $\bar{n} = 1.2 \times 10^{-4} (h/$Mpc$)^3$ 
sample at $k=0.07h$Mpc$^{-1}$. If neglected this bias would have a significant impact on cosmological parameter constraints extracted from redshift space clustering measurements which
use fitting formula or perturbation theory predictions for the dark matter velocity divergence power spectra.

Current and  future large galaxy redshift surveys will map the three-dimensional galaxy distribution to a high precision. There is an on-going major effort to advance
the models for the clustering signal in 
redshift space where the observed redshift is  composed of both the peculiar velocities of galaxies and a  cosmological redshift from the Hubble expansion.
It is well known that any scale dependent bias between halos and the dark matter would be a key systematic affecting cosmological parameter constraints.
In this paper we have used one of the highest resolution simulations to date to test currently used models for the linear bias beyond the mass limits where they were calibrated.
We also draw attention to another potentially serious systematic due to a sampling bias in the halo velocity power spectra which would affect 
the comparison between observations and any redshift space distortion model which assumes dark matter velocity statistics.
We leave further analysis  and modelling of this bias to future research.
\\
\\
\\
\\
\\
\\
\\
\section*{Acknowledgments}
The authors are grateful to Raul Angulo and Volker Springel for comments on this paper and
 for allowing the {\sc mxxl} outputs to be  used in this study.
The {\sc mxxl}
simulation was carried out on Juropa at the Juelich Supercomputer Centre in
Germany.
EJ acknowledges the support of a grant from the Simons Foundation, award number 184549. This work was supported in part by the Kavli Institute for Cosmological Physics at the University of Chicago through grants NSF PHY-0114422 and NSF PHY-0551142 and an endowment from the Kavli Foundation and its founder Fred Kavli. This work was supported by the Science and Technology Facilities Council 
[grant number ST/L00075X/1]. 
This work used the DiRAC Data Centric system at Durham University, operated by the Institute for Computational Cosmology on behalf of the STFC DiRAC HPC Facility (www.dirac.ac.uk). This equipment was funded by BIS National E-infrastructure capital grant ST/K00042X/1, STFC capital grant ST/H008519/1, and STFC DiRAC Operations grant ST/K003267/1 and Durham University. DiRAC is part of the National E-Infrastructure. 
We are grateful for the support of the University of Chicago Research Computing Center for assistance with the calculations carried out in this work.

\bibliographystyle{mn2e}
\bibliography{mybibliography}

\bsp

\label{lastpage}

\end{document}